# A Comparative Study of Hierarchical Risk Parity Portfolio and Eigen Portfolio on the NIFTY 50 Stocks

Jaydip Sen[1], Abhishek Dutta[2]

Praxis Business School, Kolkata 700104, INDIA
[1]jaydip.sen@acm.org, [2]duttaabhishek0601@gmail.com

**Abstract.** Portfolio optimization has been an area of research that has attracted a lot of attention from researchers and financial analysts. Designing an optimum portfolio is a complex task since it not only involves accurate forecasting of future stock returns and risks but also needs to optimize them. This paper presents a systematic approach to portfolio optimization using two approaches, the hierarchical risk parity algorithm and the eigen portfolio on seven sectors of the Indian stock market. The portfolios are built following the two approaches on historical stock prices from Jan 1, 2016, to Dec 31, 2020. The portfolio performances are evaluated on the test data from Jan 1, 2021, to Nov 1, 2021. The backtesting results of the portfolios indicate that the performance of the HRP portfolio is superior to that of its eigen counterpart on both training and the test data for the majority of the sectors studied.

**Keywords:** Portfolio Optimization, minimum variance portfolio, hierarchical risk parity algorithm, eigen portfolio, principal component analysis, return, risk, Sharpe ratio, prediction accuracy, backtesting.

## 1 Introduction

The design of optimized portfolios has remained a research topic of broad and intense interest among the researchers of quantitative and statistical finance for a long time. An optimum portfolio allocates the weights to a set of capital assets in a way that optimizes the return and risk of those assets. Markowitz in his seminal work proposed the mean-variance optimization approach which is based on the mean and covariance matrix of returns [1]. The algorithm, known as the *critical line algorithm* (CLA), despite the elegance in its theoretical framework, has some major limitations. One of the major problems is the adverse effects of the estimation errors in its expected returns and covariance matrix on the performance of the portfolio.

The *hierarchical risk parity* (HRP) and eigen portfolios are two well-known approaches of portfolio design that attempt to address three major shortcomings of quadratic optimization methods which are particularly relevant to the CLA [2-3]. These problems are, instability, concentration, and under-performance. Unlike the CLA, the HRP algorithm does not require the covariance matrix of the return values to be invertible. The HRP is capable of delivering good results even if the covariance matrix is ill-degenerated or singular, which is an impossibility for a quadratic opti-mizer. On the other hand, the eigen portfolio design exploits the theory of principal components to construct orthogonal components out of the stock return values and use the values of the component loading of the stocks on the first principal component as their respective weights. Interestingly, even though CLA's objective is to minimize the variance, both HRP and eigen portfolios are proven to have a higher likelihood of yielding lower out-of-sample variance than the CLA. The major weakness of the CLA algorithm is that a small deviation in the forecasted future returns can make the CLA deliver widely divergent portfolios. Given the fact that future returns cannot be forecasted with sufficient accuracy, some researchers have proposed risk-based asset allocation using the covariance matrix of the returns. However, this approach brings in another problem of instability. The instability arises because the quadratic programming methods require the inversion of a covariance matrix whose all eigenvalues must be positive. This inversion is prone to large errors when the covariance matrix is numerically ill-conditioned, i.e., when it has a high condition number [4]. The HRP and eigen portfolios are two among the new portfolio approaches that address the pitfalls of the CLA using techniques of machine learning and graph theory [2]. While HRP exploits the features of the covariance matrix without the requirement of its invertibility or positive-definiteness and works effectively on even a singular covariance matrix of returns, the eigen portfolio leverages the orthogonality properties of the principal components for explaining the variances in stock return to arrive at a better and more robust estimation of future stock returns and their volatilities [3].

Despite being recognized as two approaches that outperform the CLA algorithm, to the best of our knowledge, no study has been carried out so far to compare the performances of the HRP and the eigen portfolios on Indian stocks. This paper presents a comparative analysis of the performances of the HRP and the eigen portfolios on some important stock from selected seven sectors listed in the National Stock Exchange (NSE) of India. Based on the report of the NSE on Oct 29, 2021, the ten most significant stocks of six sectors and the 50 stocks included in the NIFTY 50 are first identified [5]. Portfolios are built using the HRP and eigen approaches for the chosen sectors using the historical prices of the stocks from Jan 1, 2016, to Dec 31, 2020. The portfolios are backtested on the in-sample data of the stock prices from Jan 1, 2016, to Dec 31, 2020, and also on the out-of-sample data of stock prices from Jan 1, 2021, to Nov 1, 2021. Extensive results on the performance of the backtesting of the portfolios are analyzed to identify the better-performing algorithm for portfolio design.

The main contribution of the current work is threefold. First, it presents two different methods of designing robust portfolios, the HRP algorithm, and the eigen approach. These portfolio design approaches are applied to seven critical sectors of stocks of the NSE. The results can be used as a guide to investors in the stock market for making profitable investments. Second, a backtesting method is proposed for evaluating the performances of the algorithms based on the daily returns yielded by the portfolios and their associated volatilities (i.e., risks). Since the backtesting is done both on the training and the test data of the stock prices, the work has identified the more efficient algorithm both on the in-sample data and the out-of-sample data. Hence, a robust framework for evaluating different portfolios is demonstrated. Third, the returns of the portfolios on the seven sectors on the test data highlight the current profitability of investment and the volatilities of the sectors studied in this work. This information can be useful for investors.

The paper is organized as follows. In Section 2, some existing works on portfolio design and stock price prediction are discussed briefly. Section 3 highlights the methodology followed. Section 4 presents the results of the two portfolio design approaches on the seven sectors. Section 5 concludes the paper.

## 2 Related Work

Due to the challenging nature of the problems and their impact on real-world applications, several propositions exist in the literature for stock price prediction and robust portfolio design for optimizing returns and risk in a portfolio. The use of predictive models based on learning algorithms and deep neural net architectures for stock price prediction is quite common [6-13]. Hybrid models are also demonstrated that combine learning-based systems with the sentiments in the unstructured non-numeric contents on the social web [14-15]. The use of multi-objective optimization, principal component analysis, and metaheuristics have also been proposed by some researchers in portfolio design [16-21]. Estimating volatility in future stock prices using GARCH has also been proposed in some work [22].

The current work presents two methods the eigen portfolio approach and the HRP method to introduce robustness while maximizing the portfolio returns for seven sectors of the NSE of India. Based on the past prices of the stocks from Jan 2016 to Dec 2020, portfolios are designed using the eigen and the HRP algorithms for each sector. The backtesting of the portfolios is carried out on the in-sample data of stock prices from Jan 2016 to Dec 31, 2020, and also on the out-of-sample data from Jan 1, 2021, to Aug 26, 2021. The backtesting is done on the return, volatility, and the Sharpe ratio of the portfolios for each sector.

## 3 Data and Methodology

In this section, the six-step approach adopted in designing the proposed system is discussed in detail. The six steps are as follows.

(1) *Choice of the sectors:* Seven important sectors of NSE are first chosen. The selected sectors are (i) auto, (ii) consumer durable, (iii) financial services, (iv) healthcare, (v) information technology, (vi) oil and gas, and (vii) NIFTY 50. NIFTY 50 contains the 50 most critical stock stocks from several sectors of the Indian stock market. For the remaining six sectors, the top ten stocks are identified based on their contributions to the computation of the overall sector index to which they belong as per the report published by the NSE on Oct 29, 2021 [5].

(2) *Data acquisition:* The prices of the ten most critical stocks of the six sectors and the 50 stocks listed in the NIFTY 50 are extracted using the *DataReader* function of the *data* sub-module of the *pandas_datareader* module in Python. The stock prices are extracted from Yahoo Finance, from Jan 1, 2016, to Nov 1, 2021. The stock price data from Jan 1, 2016, to Dec 31, 2020, are used for training the portfolios, while the portfolios are tested on the data from Jan 1, 2021, to Nov 1, 2021. Among all the features in the stock data, the variable *close* is chosen for the univariate analysis.

(3) *Derivation of the return and volatility:* The changes in the *close* values for successive days in percentage represent the *daily return*. For computing the daily returns, the *pct_change* function of Python is used. Based on the daily returns, the daily and yearly volatilities of the stocks of every sector are computed. Assuming that there are 250 operational days in a calendar year, the annual volatility values for the stocks are arrived at by multiplying the daily volatilities by a square root of 250.

(4) *Designing the eigen portfolios:* Designing eigen portfolios involves the concept of *principal component analysis* (PCA), a well-known dimensionality reduction method based on unsupervised learning. PCA retains the intrinsic variance in the data while reducing the number of dimensions. The principal components in the training data of the stock prices are determined using the PCA function defined in the *sklearn* library of Python. To retain 80% of the variance in the original stock price data, it is found that a minimum of five components is needed from the ten stocks. The components generated by the PCA function are orthogonal to each other, and their power of explanation of the variance in the data decreases with a higher component number. In other words, the first component explains the maximum percentage of the total variance. The component loading of the five principal components on each of the ten stocks reflects the weights allocated to the stocks in building the candidate eigen portfolios. Finally, the portfolio yielding the maximum Sharpe ratio among the candidates is selected as the best eigen portfolio. A Python function is used iterating over a loop for deriving the weights assigned to the five principal components and in identifying the best candidate eigen portfolio [16][21].

(5) *Hierarchical risk parity portfolio design:* As an alternative to the CLA algorithm for portfolio design, the *hierarchical risk parity* (HRP) algorithm-based portfolios are designed for the seven sectors. The HRP algorithm works in three phases: (a) *tree clustering*, (b) *quasi-diagonalization*, and (c) *recursive bisection*. These steps are briefly described in the following.

*Tree Clustering*: The tree clustering used in the HRP algorithm is an agglomerative clustering algorithm. To design the agglomerative clustering algorithm, a hierarchy class is first created in Python. The hierarchy class contains a dendrogram method that receives the value returned by a method called *linkage* defined in the same class. The linkage method received the dataset after pre-processing and transformation and computes the minimum distances between stocks based on their return values. There are several options for computing the distance. However, the *ward distance* is a good choice since it minimizes the variances in the distance between two clusters in the presence of high volatility in the stock return values. In this work, the ward distance has been used as a method to compute the distance between two clusters. The linkage method performs the clustering and returns a list of the clusters formed. The computation of linkages is followed by the visualization of the clusters through a *dendrogram*. In the dendrogram, the leaves represent the individual stocks, while the root depicts the cluster containing all the stocks. The distance between each cluster formed is represented along the *y*-axis, longer arms indicate less correlated clusters and vice versa.

*Quasi-Diagonalization*: In this step, the rows and the columns of the covariance matrix of the return values of the stocks are reorganized in such a way that the largest values lie along the diagonal. Without requiring a change in the *basis* of the covariance matrix, the quasi-diagonalization yields a very important property of the matrix – the assets (i.e., stocks) with similar return values are placed closer, while disparate assets are put at a far distance. The working principles of the algorithm are as follows. Since each row of the linkage matrix merges two branches into one, the clusters ($C_{N-1}$, 1) and ($C_{N-2}$, 2) are replaced with their constituents recursively, until there are no more clusters to merge. This recursive merging of clusters preserves the original order of the clusters [4]. The output of the algorithm is a sorted list of the original stocks (as they were before the clustering).

*Recursive Bisection*: The quasi-diagonalization step transforms the covariance matrix into a quasi-diagonal form. It is proven mathematically that allocation of weights to the assets in an inverse ratio to their variance is an optimal allocation for a *quasi-diagonal* matrix [4]. This allocation may be done in two different ways. In the bottom-up approach, the variance of a contiguous subset of stocks is computed as the variance of an inverse-variance allocation of the composite cluster. In the alternative top-down approach, the allocation among two adjacent subsets of stocks is done in inverse proportion to their aggregated variances. In the current implementation, the top-down approach is followed. A python function *computeIVP* computes the inverse-variance portfolio based on the computed variances of two clusters as its given input. The variance of a cluster is computed using another Python function called *clusterVar*. The output of the *clusterVar* function is used as the input to another Python function called *recBisect* which computes the final weights allocated to the individual stocks based on the *recursive bisection* algorithm.

The HRP algorithm performs the weight allocation to n assets in the base case in time $T(n) = O(\log_2 n)$, while its worst-case complexity is given by $T(n) = O(n)$. Unlike the MVP, which is an approximate algorithm, the HRP is an exact and deterministic algorithm. The HRP portfolios for all the sectors are built on Jan 1, 2021, based on the training data from Jan 1, 2016, to Dec 31, 2020.

(6) *Backtesting the portfolios on the training and the test data*: In the final step, the eigen and the HRP portfolios for each sector are backtested over the training and the test data set. For backtesting, the daily return values are computed for each sector for both portfolios. For the purpose of comparison, the Sharpe ratios and the aggregate volatility values for each sector are also computed for both the training and the test data. For each sector, the portfolios that perform better on the training and the test data are identified. The portfolio that performs better on the test data for a given sector is assumed to have exhibited superior performance for that sector. While it is well-known that both HRP and eigen portfolios usually outperform the CLA portfolio on the test data, a comparison of their performances on the training and the test data will be quite interesting.

## 4 Performance Evaluation

The detailed results on the performance of the portfolios and their analysis are presented in this section. The seven sectors of NSE which are selected for the analysis are as follows: (i) auto, (ii) consumer durable, (iii) financial services, (iv) healthcare, (v) IT, (vi) oil and gas, and (vii) NIFTY 50. The eigen and the HRP portfolios are implemented in Python programming language and the portfolios are trained and tested on the Google Colab platform. In the following sub-sections, the detailed results of the performances of the two portfolios on the seven sectors are presented.

## 4.1 The Auto Sector Portfolios

The ten most significant stocks and their respective contributions to the computation of the auto sector index according to the report published by the NSE on Oct 29, 2021, are as follows: Maruti Suzuki: 19.98, Mahindra & Mahindra: 15.33, Tata Motors:11.36, Bajaj Auto: 10.75, Hero MotoCorp: 7.73, Eicher Motors: 7.60, Bharat Forge: 4.18, Balkrishna Ind.: 4.15, Ashok Leyland: 4.12, and MRF: 3.58 [5].

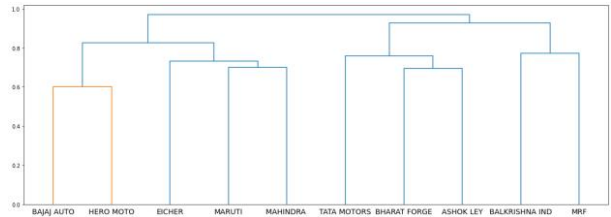

**Fig. 1.** The agglomerative clustering of the auto sector stocks – the dendrogram formed on the training data from Jan 1, 2016, to Dec 31, 2020.

The dendrogram of the clustering of the stocks of the auto sector is shown in Fig 1. The *y*-axis of the dendrogram depicts the *ward linkage* values, where a longer length of the arms signifies a higher distance, and hence less compactness in the cluster formed. For example, the cluster containing the Bajaj Auto and Hero MotoCorp stocks is the most compact one, while the one containing Balkrishna Ind. and MRF is the most heterogeneous. It is evident that the tree clustering for the HRP has created four clusters for the auto sector for the allocation of weights. These four clusters contain the following stocks: (1) Bajaj-Auto and Hero MotoCorp, (2) Eicher Motors, Maruti Suzuki, and Mahindra and Mahindra, (3) Tata Motors, Bharat Forge, and Ashok Leyland, and (4) Balkrishna Industries and MRF.

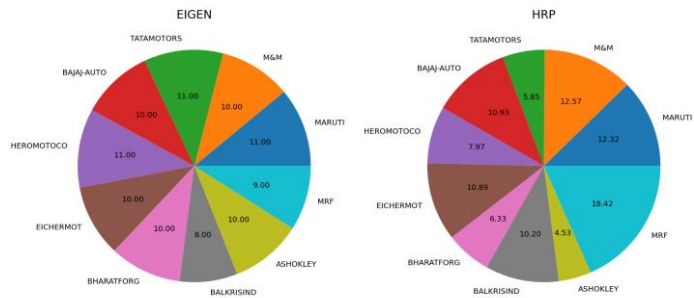

**Fig. 2.** The allocation of weights to the auto sector stocks by the EIGEN and the HRP portfolios based on stock price data from Jan 1, 2016, to Dec 31, 2020.

Fig 2 depicts the weight allocations by the eigen and the HRP portfolios for the auto sector. It is clear that both portfolios have attempted to achieve diversification in the portfolio by uniformly allocating the weights to the stocks. Fig 3(a) and Fig 3(b) show the results of backtesting for the training and the test data, respectively. These graphs plot the daily returns along the *y*-axis. The summary of the backtesting results is presented in Table 1. While the HRP portfolio is found to have produced a higher Sharpe *ratio* and lower volatility for the training (i.e., in-sample) data, the EIGEN portfolio has outperformed it on the test (i.e., out-of-sample) data producing a higher value of Sharpe ratio, albeit with slightly higher volatility.

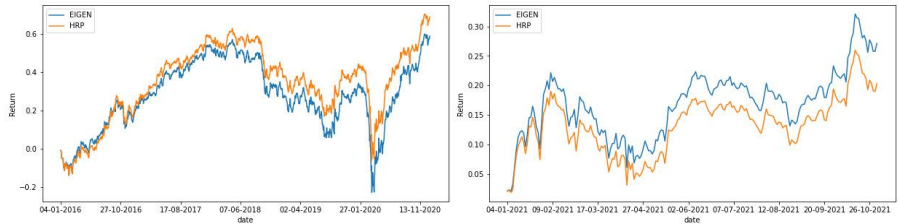

**Fig. 3.** The returns of the EIGEN and the HRP portfolios for the auto sector stocks on the (a) training data from Jan 1, 2016, to Dec 31. 2020, and (b) on the test data from Jan 1, 2021, to Nov 1, 2021.

**Table 1.** Performance of the auto sector portfolios

| Portfolio | Training Performance | | Test Performance | |
|---|---|---|---|---|
| | Volatility | Sharpe Ratio | Volatility | Sharpe Ratio |
| EIGEN | 0.240137 | 0.500069 | 0.225286 | 1.479449 |
| HRP | 0.226378 | 0.620970 | 0.207317 | 1.204434 |

### 4.2 The Consumer Durable Sector Portfolios

The top ten stocks and their respective contributions to the sectoral index of the consumer durable sector as per the NSE's report on Oct 29, 2021, are as follows: Titan Company: 35.66, Havells India: 11.54, Voltas: 10.18, Crompton Greaves Consumer Electricals: 10.03, Dixon Technologies India: 6.52, Bata India: 4.36, Kajaria Ceramics: 3.69, Relaxo Footwears: 3.50, Rajesh Exports: 3.16, Whirlpool of India: 2.56 [5]. Fig 4 depicts the clustering of the stocks by the HRP portfolio in which six clusters are apparent. These six clusters consist of the following: (i) Crompton and Dixon, (ii) Relaxo, (iii) Titan and Britannia, (iv) Havells and Voltas, (v) Whirlpool and Kajaria, and (v) Rajesh Exports. Fig 5 shows the weight allocation by the two portfolios, while the returns on the training and test data are shown in Fig 6. The Sharpe ratio of the HRP portfolio is higher for the test data as observed in Table 2.

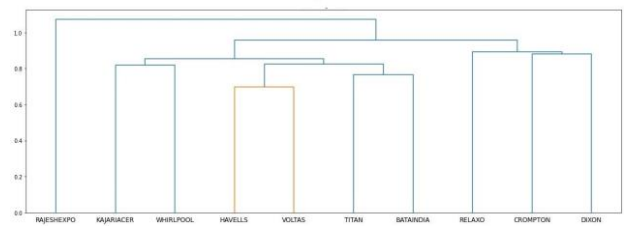

**Fig. 4.** The agglomerative clustering of the consumer durable sector stocks – the dendrogram formed on the training data from Jan 1, 2016, to Dec 31, 2020.

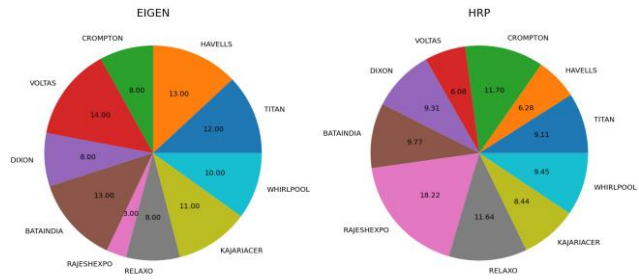

**Fig. 5.** The allocation of weights to the consumer durable sector stocks by the EIGEN and the HRP portfolios based on stock price data from Jan 1, 2016, to Dec 31, 2020.

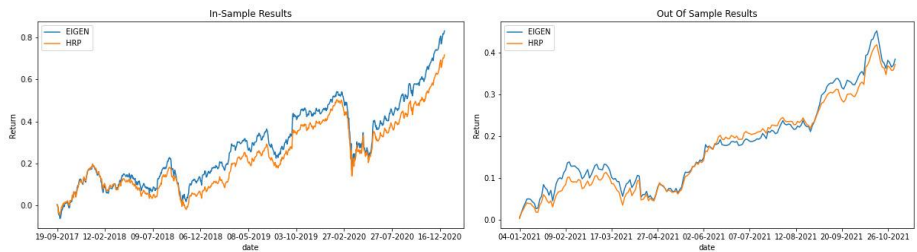

**Fig. 6.** The returns of the EIGEN and the HRP portfolios for the consumer durable sector stocks on the (a) training data from Jan 1, 2016, to Dec 31. 2020, and (b) on the test data from Jan 1, 2021, to Nov 1, 2021.

**Table 2.** Performance of the consumer durable sector portfolios

| Portfolio | Training Performance | | Test Performance | |
|---|---|---|---|---|
| | Volatility | Sharpe Ratio | Volatility | Sharpe Ratio |
| EIGEN | 0.205992 | 1.253044 | 0.172882 | 2.731261 |
| HRP | 0.184684 | 1.205891 | 0.151220 | 3.019343 |

### 4.3 The Financial Services Sector Portfolios

The top ten stocks and their respective contributions to the sectoral index of the financial services sector as per the NSE's report on Oct 29, 2021, are as follows: HDFC Bank: 22.49, ICICI Bank: 18.06, Housing Development Finance Corporation: 16.68, Kotak Mahindra Bank: 9.68, Bajaj Finance: 6.38, State Bank of India: 6.26, Axis Bank: 6.21, Bajaj Finserv: 3.50, HDFC Life Insurance Company: 2.06, and SBI Life Insurance Company: 1.64 [5]. The clustering done by the HRP on the stocks of this sector is shown in Fig 7, in which the clusters consist of the following stocks: (i) Axis Bank, (ii) ICICI Bank and SBI, (iii) Bajaj Fin and Bajaj Finserv, (iv) Kotak Bank, and (v) HDFC Bank and HDFC Bank. The allocation of the weights by the portfolios is shown in Fig 8, while the returns yielded are shown in Fig 9. The results presented in Table 3 indicate that the EIGEN portfolio has yielded a higher Sharpe ratio for the test data. However, on the training data, the HRP portfolio produced a higher Sharpe ratio. The volatilities of the HRP portfolios are lower for both datasets.

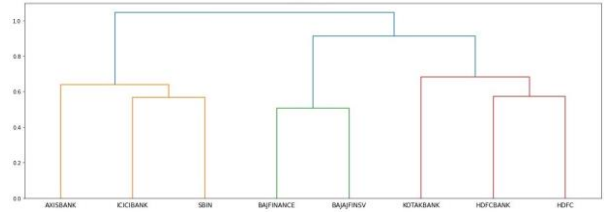

**Fig. 7.** The agglomerative clustering of the financial services sector stocks – the dendrogram formed on the training data from Jan 1, 2016, to Dec 31, 2020.

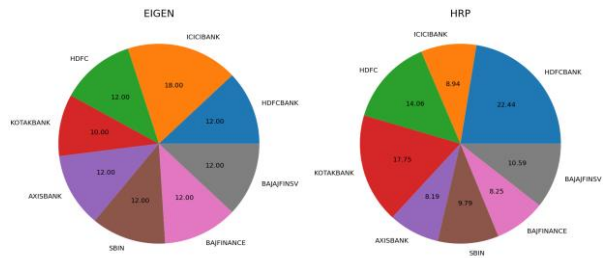

**Fig. 8.** The allocation of weights to the financial services sector stocks by the EIGEN and the HRP portfolios based on stock price data from Jan 1, 2016, to Dec 31, 2020.

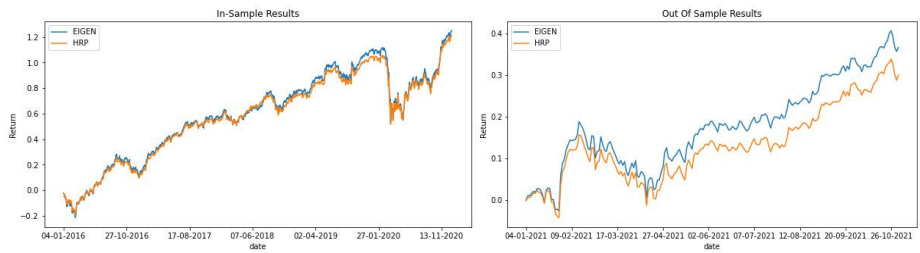

**Fig. 9.** The returns of the EIGEN and the HRP portfolios for the financial services sector stocks on the (a) training data from Jan 1, 2016, to Dec 31. 2020, and (b) on the test data from Jan 1, 2021, to Nov 1, 2021.

### 4.4 The Healthcare Sector Portfolios

The top ten stocks and their respective contributions to the sectoral index of the healthcare sector as per the NSE's report on Oct 29, 2021, are as follows: Sun Pharmaceutical Industries: 17.88, Divi's Laboratories: 13.67, Dr. Reddy's Laboratories: 11.79, Cipla: 9.58, Apollo Hospitals Enterprise: 8.94, Lupin: 4.63, Laurus Labs: 4.21, Aurobindo Pharma: 4.04, Alkem Laboratories: 3.51, and Biocon: 3.34 [5]. Fig. 10 shows the dendrogram of the clustering done by the HRP portfolio, in which seven clusters are visible. The weights allocation and the returns yielded by the two portfolios are presented in Fig. 11 and Fig. 12, respectively. The results in Table 4 show that the Sharpe ratios produced by the HRP portfolio are higher in both cases.

**Table 3.** Performance of the financial services sector portfolios

| Portfolio | Training Performance | | Test Performance | |
|---|---|---|---|---|
| | Volatility | Sharpe Ratio | Volatility | Sharpe Ratio |
| EIGEN | 0.262314 | 0.973574 | 0.236104 | 1.908093 |
| HRP | 0.242130 | 1.024041 | 0.225178 | 1.637399 |

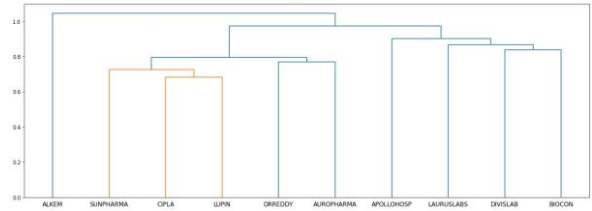

**Fig. 10.** The agglomerative clustering of the healthcare sector stocks – the dendrogram formed on the training data from Jan 1, 2016, to Dec 31, 2020.

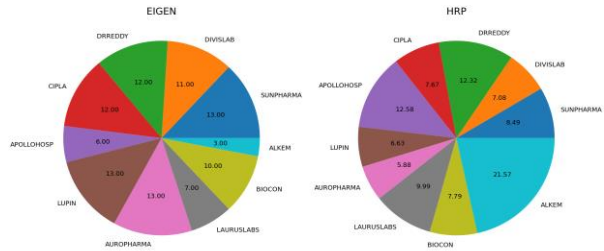

**Fig. 11.** The allocation of weights to the healthcare sector stocks by the EIGEN and the HRP portfolios based on stock price data from Jan 1, 2016, to Dec 31, 2020.

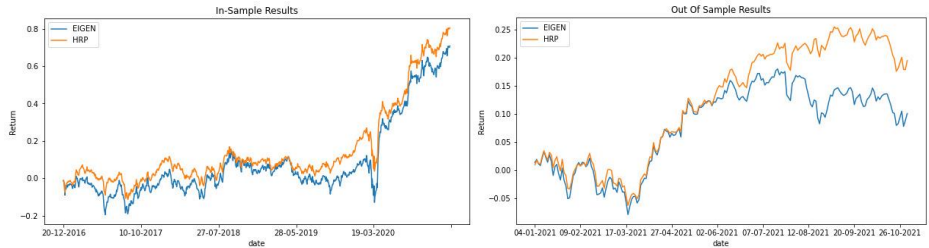

**Fig. 12.** The returns of the EIGEN and the HRP portfolios for the healthcare services sector stocks on the (a) training data from Jan 1, 2016, to Dec 31. 2020, and (b) on the test data from Jan 1, 2021, to Nov 1, 2021.

**Table 4.** Performance of the healthcare sector portfolios

| Portfolio | Training Performance | | Test Performance | |
|---|---|---|---|---|
| | Volatility | Sharpe Ratio | Volatility | Sharpe Ratio |
| EIGEN | 0.223730 | 0.799622 | 0.184003 | 0.672148 |
| HRP | 0.193036 | 1.054018 | 0.169768 | 1.410061 |

## 4.5 The Information Technology Sector Portfolios

The top ten stocks and their respective contributions to the sectoral index of the media sector as per the NSE's report on Oct 29, 2021, are as follows: Infosys: 27.21, Tata Consultancy Services: 24.05, Tech Mahindra: 9.70, Wipro: 9.50, HCL Technologies: 8.48, Larsen & Toubro Infotech: 5.74, MindTree: 5.45, MphasiS: 5.03, L&T Technology Services: 2.44, Coverage: 2.40 [5]. Fig. 13 depicts the dendrogram of clustering by the HRP portfolio in which seven clusters are visible. Fig. 14 and Fig. 15 show the weight allocation and the returns yielded by the two portfolios on the information technology sector stocks, respectively. It is observed from Table 5 that the HRP portfolio has yielded a higher Sharpe ratio on the test data.

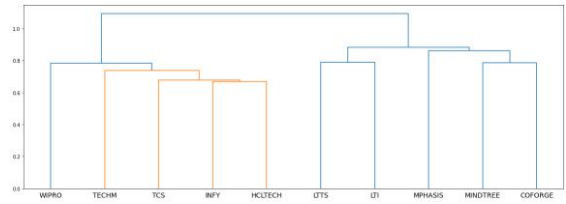

**Fig. 13.** The agglomerative clustering of the information technology sector stocks – the dendrogram formed on the training data from Jan 1, 2016, to Dec 31, 2020.

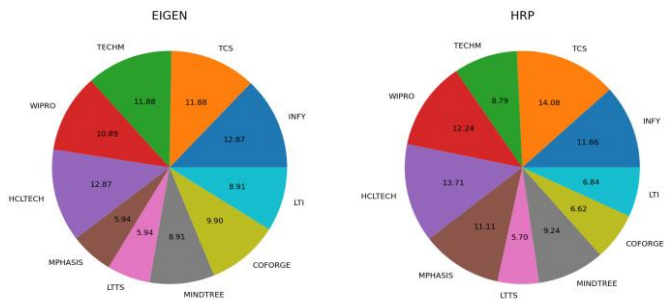

**Fig. 14.** The allocation of weights to the media sector stocks by the EIGEN and the HRP portfolios based on stock price data from Jan 1, 2016, to Dec 31, 2020.

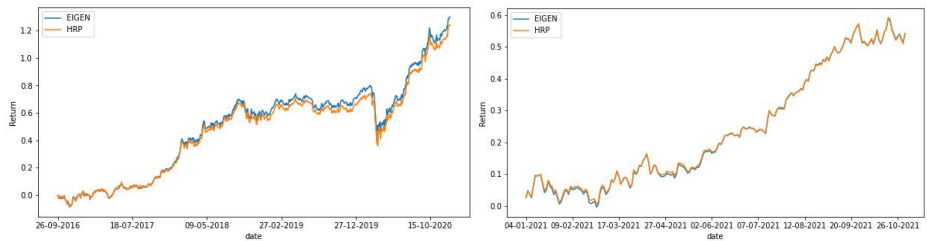

**Fig. 15.** The returns of the EIGEN and the HRP portfolios for the information technology sector stocks on the (a) the training data from Jan 1, 2016, to Dec 31. 2020, and (b) on the test data from Jan 1, 2021, to Nov 1, 2021.

**Table 5.** Performance of the information technology sector portfolios

| Portfolio | Training Performance | | Test Performance | |
|---|---|---|---|---|
| | Volatility | Sharpe Ratio | Volatility | Sharpe Ratio |
| EIGEN | 0.214254 | 1.44903 | 0.234554 | 2.839298 |
| HRP | 0.206345 | 1.43674 | 0.227796 | 2.910402 |

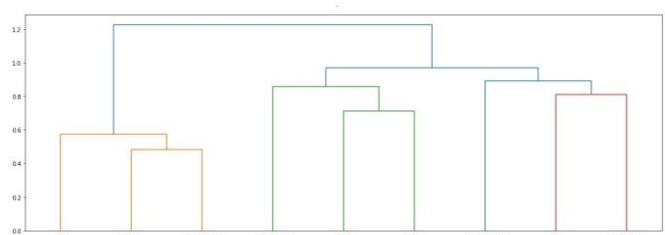

**Fig. 16.** The agglomerative clustering of the oil and gas sector stocks – the dendrogram formed on the training data from Jan 1, 2016, to Dec 31, 2020.

### 4.6 The Oil and Gas Sector Portfolios

The top ten stocks and their respective contributions to the sectoral index of the oil and gas sector as per the NSE's report on Oct 29, 2021, are as follows: Reliance Industries: 32.78, Oil & Natural Gas Corporation: 12.70, Bharat Petroleum Corporation: 9.31, Adani Total Gas: 9.24, Indian Oil Corporation: 7.60, GAIL India: 6.33, Hindustan Petroleum Corporation: 4.63, Petronet LNG: 4.02, Indraprastha Gas: 3.87, and Gujarat Gas: 2.50 [5]. The HRP portfolio has created six clusters on the ten stocks of this sector as observed in Fig. 16. The weight allocation by the portfolios and their returns are depicted in Fig. 17 and Fig. 18, respectively. The results presented in Table 6 show that the Sharpe ratio yielded by the HRP portfolio is higher for both training and test data. Moreover, for both cases, the volatilities of the HRP portfolio are found to be lower.

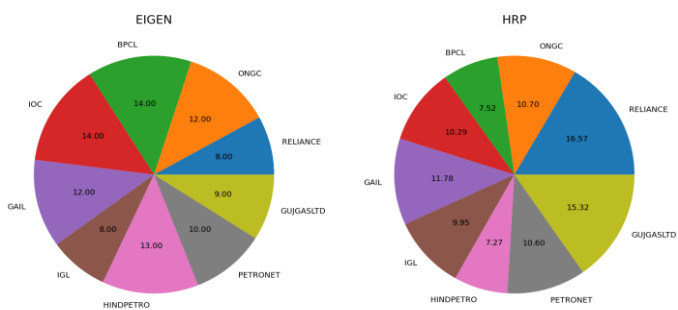

**Fig. 17.** The allocation of weights to the oil and gas sector stocks by the EIGEN and the HRP portfolios based on stock price data from Jan 1, 2016, to Dec 31, 2020.

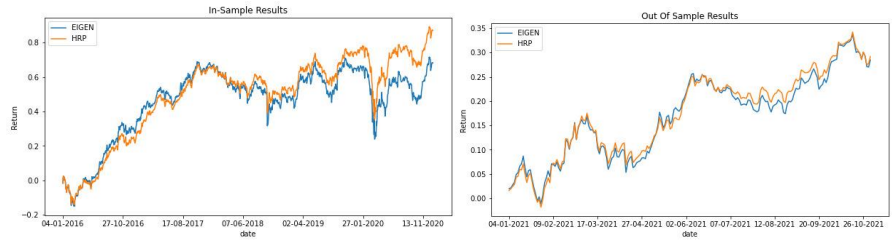

**Fig. 18.** The returns of the EIGEN and the HRP portfolios for the oil and gas sector stocks on the (a) training data from Jan 1, 2016, to Dec 31. 2020, and (b) on the test data from Jan 1, 2021, to Nov 1, 2021.

**Table 6.** Performance of the oil and gas sector portfolios

| Portfolio | Training Performance | | Test Performance | |
|---|---|---|---|---|
| | Volatility | Sharpe Ratio | Volatility | Sharpe Ratio |
| EIGEN | 0.236194 | 0.590050 | 0.203017 | 1.723656 |
| HRP | 0.213704 | 0.832144 | 0.193843 | 1.848685 |

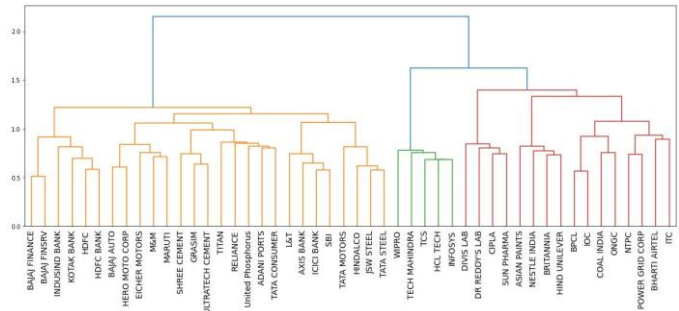

**Fig. 19.** The agglomerative clustering of the NIFTY 50 stocks – the dendrogram formed on the training data from Jan 1, 2016, to Dec 31, 2020.

### 4.7 The NIFTY 50 Portfolios

Finally, we consider the NIFTY 50 stocks and construct the EIGEN and the HRP portfolios for them. These stocks are the market leaders across 13 sectors in the NSE and have a low-risk quotient [5]. The dendrogram is depicted in Fig. 19, in which three distinct clusters are visible. The allocation of the weights and the returns yielded by the portfolios are shown in Fig 20 and Fig 21, respectively. The allocations made by both portfolios appear to be very uniform indicating diversified portfolios. The results presented in Table 7 show that the HRP portfolio has outperformed its EIGEN counterpart as it has yielded a higher value of the Sharpe ratio.

In summary, it is observed that among the seven sectors, the HRP portfolio has produced higher Sharpe ratios for four sectors on the training data, while it outperformed the EIGEN portfolio on five sectors on the test data. The results show that the performance of the HRP portfolio is superior to that of the EIGEN on what matters most, i.e., the test data of the sectors. Even if it is not right to generalize this observation found in the study of the seven sectors, it appears that the diversification ap-proach of HRP is more effective than that of the eigen portfolio approach. In order words, in the presence of some highly correlated stocks that lead to a highly unstable inverse correlation matrix, the HRP's approach of diversification by assigning weights to stocks in the same cluster based on the reciprocals of their variances seems to be more effective than the formation of principal components and using their loading values as the portfolio weights.

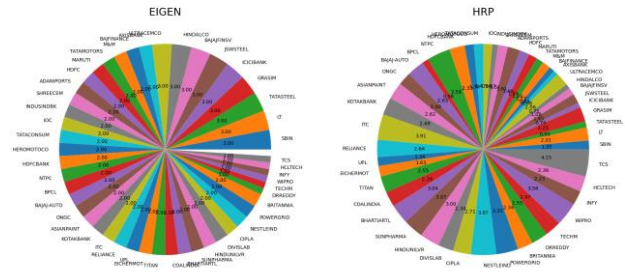

**Fig. 20.** The allocation of weights to the NIFTY 50 stocks by the EIGEN and the HRP portfolios based on stock price data from Jan 1, 2016, to Dec 31, 2020.

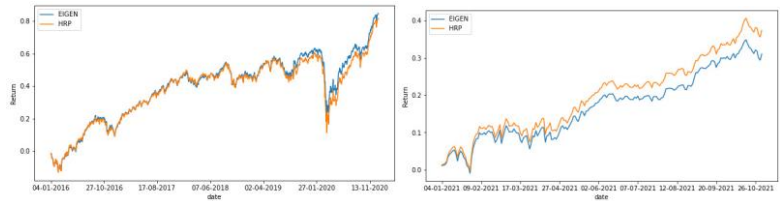

**Fig. 21.** The returns of the EIGEN and the HRP portfolios for the NIFTY 50 stocks on the test data from Jan 1, 2021, to Aug 26, 2021.

Table 7. Performance of the NIFTY 50 portfolios

| Portfolio | Training Performance | | Test Performance | |
|---|---|---|---|---|
| | Volatility | Sharpe Ratio | Volatility | Sharpe Ratio |
| EIGEN | 0.184898 | 0.934880 | 0.153761 | 2.480044 |
| HRP | 0.187925 | 0.887088 | 0.163927 | 2.799373 |

## 4  Conclusion

This paper has presented portfolio design approaches on some important sectors of the Indian stock market using the EIGEN and HRP algorithms. Based on the past prices of the top ten stocks of six sectors and the 50 stocks from NIFTY 50, the portfolios are designed. While the stock price data from Jan 1, 2016, to Dec 31, 2020, are used for building the portfolios, the period Jan 1, 2021, to Nov 1, 2021, is used for the testing. It is found that in both the training and the test data, the HRP portfolio yielded higher Sharpe ratios for the majority of the sectors among the seven sectors studied in this work. It is evident that the HRP portfolio has not only learned more effectively from the patterns in the training data but also achieved better diversification leading to higher Sharpe ratios on the test data.

## References

1. Markowitz, H.: Portfolio Selection. Journal of Finance, 7(1), 77-91 (1952)
2. de Prado, M. L.: Building Diversified Portfolios that Outperform Out of Sample. Journal of Portfolio Management, 42(4), 59-69 (2016)
3. Peng, Y., Albuquerque, P. H. M, Do Nascimento, I-F., and Machado, J. V. F.: Between Nonlinearities, Complexity, and Noises: An Application on Portfolio Selection Using Kernel Principal Component Analysis. Entropy, 21(4), 376 (2019)
4. Baily, D. and de Prado, M. L.: Balanced Baskets: A New Approach to Trading and Hedging Risks. Journal of Investment Strategies, 1(4), 61-62 (2012)
5. NSE Website: http://www1.nseindia.com (Accessed on Jan 03, 2021)
6. Mehtab, S., Sen, J., and Dutta, A.: Stock Price Prediction Using Machine Learning and LSTM-Based Deep Learning Model. In: Proc. of SoMMA, pp. 88-106 (2020)
7. Mehtab, S. and Sen, J.: Stock Price Prediction Using Convolutional Neural Networks on a Multivariate Time Series. In: Proc. of 2nd NCMLAI, New Delhi, India (2020)
8. Sen, J.: Stock Price Prediction Using Machine Learning and Deep Learning Frameworks. In: Proc. of ICBAI, Bangalore, India (2018)
9. Sen, J. and Datta Chaudhuri, T.: A Robust Predictive Model for Stock Price Forecasting. In: Proc. of ICBAI, Bangalore, India (2017)
10. Sen, J. and Mehtab, S.: Accurate Stock Price Forecasting Using Robust and Optimized Deep Learning Model. In: Proc. of IEEE CONIT, Hubli, India (2021)
11. Mehtab, S. and Sen, J.: Stock Price Prediction Using CNN and LSTM-Based Deep Learning Models. In: Proc. of IEEE DASA, Bahrain (2020)
12. Qiu, J. and Wang, B.: Forecasting Stock Prices with Long-Short Term Memory Neural Network Based on Attention Mechanism. PLOS ONE, 15(1), e0227222 (2020)
13. Mehtab, S. and Sen, J.: A Time Series Analysis-Based Stock Price Prediction Using Machine Learning and Deep Learning Models. Int. J. of Business Forecasting and Marketing Intelligence, 6(4), 272-335 (2020)
14. Mehtab, S. and Sen, J.: A Robust Predictive Model for Stock Price Prediction Using Deep Learning and Natural Language Processing. In: Proc. of 7th BAICONF (2019)
15. Carta, S. M., Consoli, S., Piras, L., Podda, S., and Recupero, D. R.: Explainable Machine Learning Exploiting News and Domain-Specific Lexicon for Stock Market Forecasting. IEEE Access, 9, 30193-302015 (2021)
16. Sen, J. and Mehtab, S.: A Comparative Study of Optimum Risk Portfolio and Eigen Portfolio on the Indian Stock Market. Int. J. of Business Forecasting and Marketing Intelligence, 7(2), 143-193, Inderscience Publishers, in press (2021)
17. Corazza, M., Di Tollo, G., Fasano, G., and Pesenti, R.: A Novel Hybrid PSO-Based Metaheuristic for Costly Portfolio Selection Problem. Annals of Operations Research, 304, 109-137 (2021)
18. Sen, J., Dutta, A., and Mehtab, S.: Stock Portfolio Optimization Using a Deep Learning LSTM Model. In: Proc. of IEEE MysuruCon, Hassan, India (2021)
19. Sen, J., Mondal, S., and Mehtab, S.: Portfolio Optimization on NIFTY Thematic Sector Stocks Using an LSTM Model. In: Proc. of IEEE ICDABI, Bahrain (2021)
20. Wang, Z., Zhang, X., Zhang, Z., and Sheng, D.: Credit Portfolio Optimization: A Multi-Objective Genetic Algorithm Approach. Bora Istanbul Review, in press (2021)
21. Erwin, K. and Engelbrecht, A.: Improved set-based particle swarm optimization for portfolio optimization. In: Proc. of IEEE SSCI, Canberra, Australia (2020)
22. Sen, J., Mehtab, S., and Dutta, A.: Volatility Modeling of Stock from Selected Sectors of the Indian Economy Using GARCH. In: Proc. of IEEE ASIANCON, Pune, India (2021)